\documentclass[conference]{IEEEtran}

\IEEEoverridecommandlockouts 
\usepackage{booktabs}
\usepackage{textcomp}     

\usepackage{amsmath}
\usepackage{amssymb}
\usepackage{xcolor}
\usepackage{graphicx}
\usepackage[font=small]{caption}
\usepackage[subrefformat=parens]{subcaption}
\usepackage{amsthm}
\usepackage{enumitem}
\usepackage{float}
\usepackage{bm}
\usepackage{enumitem}
\usepackage[noend]{algorithmic}
\usepackage{algorithm}
\usepackage{multirow}
\usepackage{soul}
\usepackage{cite}
\usepackage{mathtools}
\usepackage{xspace}

\theoremstyle{definition}

\newtheorem{definition}{Definition}
\usepackage{tikz}
\usepackage{pgfplots}
\pgfplotsset{compat=1.18}
\usetikzlibrary{plotmarks}
\usetikzlibrary{matrix,positioning,shapes,arrows,decorations,calc,fit}
\usetikzlibrary{decorations.pathreplacing}
\usetikzlibrary{spy,backgrounds}
\usetikzlibrary{external}
\usetikzlibrary{patterns}
\usetikzlibrary{shapes,arrows}
\usetikzlibrary{shadows.blur}
\usetikzlibrary{shapes.symbols}
\usetikzlibrary{
    pgfplots.groupplots,
    matrix
}

\usetikzlibrary{arrows.meta,positioning,fit,calc}

\usetikzlibrary{calc, arrows.meta}

\IEEEaftertitletext{\vspace{-0.5\baselineskip}}
\newcommand{\Ftwo}{\mathbb{F}_2}
\newcommand{\Feight}{\mathbb{F}_{2^8}}

\newcommand{\Prob}{\mathbb{P}}
\newcommand{\Pc}{\mathbb{P}^c}
\newcommand{\Ps}{\mathbb{P}^s}
\newcommand{\Lc}{L^c}
\newcommand{\Ls}{L^s}
\newcommand{\tLc}{\tilde L^c}
\newcommand{\tLs}{\tilde L^s}

\newcommand{\lc}{\lambda^c}
\newcommand{\ls}{\lambda^s}
\newcommand{\tlc}{\tilde\lambda^c}
\newcommand{\tls}{\tilde\lambda^s}
\newcommand{\lf}{\lambda}
\newcommand{\Ntep}{N_{\mathrm{TEP}}}
\newcommand{\Tbit}{\mathcal{T}_b}
\newcommand{\Tbyte}{\mathcal{T}_B}
\newcommand{\semOSD}{Sem-OSD\xspace}

\begin{document}

\title{Semantic Ordered Statistics Decoding}

\author{Chentao Yue, Branka Vucetic, and Yonghui Li%
\thanks{The work of Chentao Yue was supported by ARC DECRA under Grant DE250101332. Code available: https://github.com/ChentaoYue/sem-OSD.}%
\\
School of Electrical and Computer Engineering, The University of Sydney, Sydney, NSW 2006, Australia\\
E-mail: \{chentao.yue, branka.vucetic, yonghui.li\}@sydney.edu.au
}

\maketitle

\begin{abstract}
We propose a Semantic Ordered Statistics Decoder (\semOSD), a soft decoder for short linear block codes {carrying byte-streamed sources such as natural-language text.}
{\semOSD\ injects a byte-level language-model (LM) prior into ordered statistics decoding (OSD) through a fused bit-level score that combines channel reliability with the LM prior, and uses it for the most-reliable basis (MRB) selection and the codeword candidate scoring. \semOSD\ enumerates two complementary test-error-pattern (TEP) families: a bit-flip family that flips up to $m$ bits, and an LM-driven family of up to $\omega$ byte substitutions that reaches error patterns the bit-flip family cannot.} The LM prior is computed by a byte-level Transformer fine-tuned for byte-level denoising.
{Simulation results show that, on AWGN, \semOSD\ achieves block error rates (BLERs) below the finite-blocklength normal-approximation bound for uniform sources on both binary BCH$(127,64)$ and shortened RS$(16,8)$ over $\Feight$, exceeding Fossorier OSD by a $1.5$\,dB coding gain. On a Gilbert--Elliott burst-error channel, \semOSD\ provides $4$\,dB and $1$\,dB of more coding gain than Berlekamp--Massey and OSD, respectively.}
\end{abstract}
\vspace{-0.75em}

\vspace{-0.5em}
\section{Introduction}\label{sec:intro}
\vspace{-0.5em}

Modern wireless networks carry traffic whose statistical structure is far from random: natural-language text, speech and video streams, structured records, and the prompts and responses exchanged with large language models. {These sources exhibit strong inner correlations across symbols, commonly understood as \emph{semantics}, which pretrained generative language models can quantify as a byte or token-level distribution.}

Classical channel decoders nevertheless ignore this structure. {Conventional decoders, including} Berlekamp--Massey (BM) decoding of Reed--Solomon (RS) and Bose--Chaudhuri--Hocquenghem (BCH) codes~\cite{berlekamp1968algebraic}, belief propagation (BP) on low-density parity-check (LDPC) codes, and Viterbi decoding of convolutional codes, all treat the information word as drawn uniformly over its alphabet. {Soft-decision decoding refines decoding error rate through richer candidate enumeration: Chase decoding~\cite{chase1972} {enumerates} candidate codewords, while} ordered statistics decoding (OSD)~\cite{fossorier1995soft} and its complexity-reduced variants~{\cite{jiang2008,yue2021probability}} enumerate test-error patterns {(TEPs)} inside a most-reliable basis {(MRB)} derived from channel reliability. All make the same uniform-source assumption, and exploit no source structure.

A separate paradigm replaces the algebraic channel coding entirely. Semantic communication trains an end-to-end neural transceiver as joint source--channel coding (JSCC)~{\cite{gunduz2023beyond,bourtsoulatze2019deep}}; recent variants embed large language models (LLMs) into the transceiver as semantic encoders or decoder-side knowledge~{\cite{jiang2024large,nam2024language}}, using the LLM both for source representation and for learned error correction. These schemes capture source structure but give up the error rate and complexity guarantees of a classical channel code.

{Several recent works keep the algebraic code and use a language model (LM) as a decoder-side prior. Hao~\emph{et al.}~\cite{hao2025short} decode each eight-byte BCH block independently and apply an LM corrector to the hard-decision output, with no access to channel soft information. Wang~\emph{et al.}~\cite{clsec} combine post-decoding bit {log-likelihood ratios (LLRs)} with an LM word-level posterior via a Bayesian product, but the combination acts only on words flagged as erroneous after channel decoding, so neither layer refines the decoder's search itself. Li~\emph{et al.}~\cite{llmviterbi} use an LM to re-rank Viterbi survivor paths, yet the paths themselves are generated by Viterbi alone. In these approaches, the LM refines the channel decoder's outputs rather than generating new codeword candidates.}


{This paper focuses on byte-level natural-language text transmission, whose source entropy is well below the random bit distribution~\cite{shannon1951prediction}. The distribution of natural language is accessible through pretrained byte-level language models. We inject this LM prior into OSD~\cite{fossorier1995soft}, a universal near-maximum-likelihood decoder (MLD) for short linear block codes. The framework generalizes to other source coding and models.}

{We propose Semantic Ordered Statistics Decoding (\semOSD), an LM-aided OSD for streaming sources, where prior decoded blocks supply a clean linguistic context for the current block. \semOSD fuses the byte-level LM prior with the channel reliability into a single bit-level score. The score drives both the MRB selection and the candidate scoring. \semOSD\ enumerates two complementary TEP families: (i) a bit-flip family that flips up to $m$ bits, inherited from Fossorier OSD and suited to isolated bit errors; and (ii) an LM-driven family of up to $\omega$ byte substitutions, suited to burst errors that the bit-flip family cannot reach. The joint search over the two families covers isolated-bit and burst-error regimes within a single decoder.} We instantiate \semOSD\ on shortened RS$(16,8)$ over $\Feight$ {and} binary BCH$(127,64)$.  On AWGN at low-to-moderate signal-to-noise ratio (SNR), \semOSD\ {achieves block error rate (BLER) below the finite-blocklength normal-approximation bound~\cite{PPV2010l} for uniform sources, with an order-of-magnitude reduction over Fossorier OSD}. On a Gilbert--Elliott channel under state-unaware block-average reception, {\semOSD\ delivers $4$\,dB of coding gain over BM and $1$\,dB over Fossorier OSD}.

The paper is organized as follows. Section~\ref{sec:prelim} provides preliminaries, Sections~\ref{sec:priors} and \ref{sec:semosd} develop the algorithm, Section~\ref{sec:exp} reports simulation results, and Section~\ref{sec:conclusion} concludes.

\section{Preliminaries}\label{sec:prelim}
\vspace{-0.3em}


\subsection{System Model}\label{sec:prelim:system}
\vspace{-0.3em}

{We consider a stream of natural-language text segmented into sentences, each partitioned into $G$ consecutive $k$-byte blocks.} The $g$-th block of a sentence is the byte vector $\boldsymbol{\mu}^{(g)}\in\Feight^k$, $g\in\{0,\ldots,G\!-\!1\}$. Each block is encoded into a bit-level codeword $\mathbf{c}^{(g)} = (c_0^{(g)},\ldots,c_{n_b-1}^{(g)}) \in \Ftwo^{n_b}$ (see Section~\ref{sec:prelim:codes}), BPSK-modulated to $s_\ell^{(g)} = 1 - 2c_\ell^{(g)} \in \{\pm 1\}$, and transmitted over the channel of Section~\ref{sec:prelim:channels}. The receiver observes
\begin{equation}\label{eq:obs}
y^{(g)}_\ell = s^{(g)}_\ell + z^{(g)}_\ell, \qquad \ell=0,\ldots,n_b-1,
\end{equation}
with $z^{(g)}_\ell\sim\mathcal{N}(0,\sigma^2_{\mathrm{eff},\ell})$, where $\sigma^2_{\mathrm{eff},\ell}=\sigma^2$ on AWGN and $\sigma^2_{\mathrm{eff},\ell}=\sigma^2_{S^{(g)}_\ell}$ on Gilbert--Elliott.

\semOSD\ decodes one block at a time and uses the bytes of the previous blocks of the same sentence as a clean linguistic context for the current block. When decoding block $g$, the decoder is provided with the prefix
\begin{equation}\label{eq:ctx}
\mathrm{ctx}^{(g)} = (\boldsymbol{\mu}^{(0)},\ldots,\boldsymbol{\mu}^{(g-1)})
\end{equation}
in error-free form, for any  $g> 0$. {This assumption is consistent with the streaming setting: prior blocks of the same sentence have already been acknowledged through hybrid-ARQ or successfully decoded by the time block $g$ arrives, so the prefix is reliably available at the receiver. Block $g$ itself is decoded under the channel model of~\eqref{eq:obs}.} The remainder of the paper analyses one such block and drops the index $(g)$.

\vspace{-0.3em}
\subsection{Channels and Reception}\label{sec:prelim:channels}
\vspace{-0.3em}
We evaluate on two channels at matched $E_b/N_0$, where $E_b$ is the energy per information bit and {$N_0$ is the one-sided noise spectral density}. The AWGN channel adds an independent Gaussian sample of variance $\sigma^2=1/(2R\gamma_b)$ to each transmitted bit, with $\gamma_b=10^{E_b/N_0/10}$ and {$R=k_b/n_b$ the code rate}. The Gilbert--Elliott channel~\cite{gilbert1960capacity,elliott1963estimates} models bursty {noise} through a two-state Markov chain {$S_\ell\in\{G,B\}$, stepping once per transmitted bit within a block,} with stationary bad-state probability $\pi_B$ and mean burst length $\bar L=1/p_{BG}$. Conditioned on {$S_\ell$}, the bit is observed in AWGN of variance $\sigma_G^2$ if $S_\ell=G$ and $\sigma_B^2$ if $S_\ell=B$, with $\rho^2\triangleq\sigma_B^2/\sigma_G^2\gg 1$. The pair $(\sigma_G^2,\sigma_B^2)$ is calibrated so that
\begin{equation}\label{eq:siginst}
\hat\sigma^2 \;\triangleq\; (1-\pi_B)\,\sigma_G^2+\pi_B\,\sigma_B^2 \;=\; \frac{1}{2R\,\gamma_b},
\end{equation}
placing the AWGN and Gilbert--Elliott curves on a common $E_b/N_0$ axis: both channels deliver the same total noise energy per bit, but GE concentrates that energy into infrequent bursts.

We assume that the receiver is \emph{state-unaware} in Gilbert--Elliott channel. It forms bit LLR $L_\ell=2y_\ell/\hat\sigma^2$ from the block-average variance and never observes $S_\ell$.

\vspace{-0.3em}
\subsection{Linear Codes and Ordered Statistics Decoding}\label{sec:prelim:codes}
\vspace{-0.3em}
We work with linear block codes of length $n_b$ bits and information length $k_b=8k$ bits, encoded by a binary generator matrix $\mathbf{G}_b\in\Ftwo^{k_b\times n_b}$ in systematic form $[\mathbf{I}_{k_b}\mid\mathbf{P}]$, with minimum Hamming distance $d_{\min}$. The byte-alignment $k_b=8k$ {lets each information byte $\mu_i\in\Feight$ expand bit-wise as $u_{8i+j}=\mathrm{bit}_j(\mu_i)$ for $j=0,\ldots,7$, where $\mathrm{bit}_j(\nu)\in\Ftwo$ denotes the $j$-th bit of $\nu$; the bit vector $\mathbf{u}\in\Ftwo^{k_b}$ is encoded as $\mathbf{c}=\mathbf{u}\mathbf{G}_b\in\Ftwo^{n_b}$}. We instantiate \semOSD\ on two short codes with $k_b=64$. BCH$(127,64)$ is the \emph{binary} case, with the standard generator polynomial~\cite{bose1960class}. Reed--Solomon RS$(16,8)$~\cite{reed1960polynomial} is the \emph{nonbinary} case, defined over $\Feight$ so each codeword symbol carries one byte. RS codes are naturally suited to burst channels, as a burst that flips several consecutive bits is absorbed as a single symbol error.

OSD~\cite{fossorier1995soft} is a universal soft decoder for any binary linear block code, approaching MLD as the order parameter $m$ increases. Order-$m$ OSD proceeds in four steps: (i) compute a reliability $r_\ell$ for every received bit, typically the magnitude of the bit LLR, and form the permutation $\pi$ that sorts $\{r_\ell\}$ in decreasing order; (ii) apply $\pi$ to the columns of $\mathbf{G}_b$ and Gaussian eliminate over $\Ftwo$ to obtain a systematic generator $\mathbf{G}_s=[\mathbf{I}_{k_b}\mid\mathbf{P}']$, swapping columns for rank if needed, so that the leading $k_b$ permuted positions form the MRB; (iii) for each TEP $\mathbf{e}\in\Ftwo^{k_b}$ of Hamming weight at most $m$, re-encode $(\mathbf{u}_0\oplus\mathbf{e})\mathbf{G}_s$, where $\mathbf{u}_0$ is the bit-level hard decision at the MRB; (iv) select the candidate that maximises a soft score against the channel observation, and un-permute it. 

OSD requires $\sum_{w=0}^m\binom{k_b}{w}$ TEP evaluations, which grows steeply with $m$. \semOSD\ retains this skeleton and modifies three components, the reliability metric $r_\ell$, the TEP set, and the candidate score, as developed in Section~\ref{sec:semosd}.

\vspace{-0.8em}
\section{Channel and Semantic Priors}\label{sec:priors}
\vspace{-0.3em}

\begin{figure}[t]
\centering
\begin{tikzpicture}[
  >=Latex, font=\scriptsize,
  block/.style={draw, rounded corners=2pt, thick, align=center,
                inner xsep=4pt, inner ysep=2.5pt, minimum height=5mm}
]
\node[block, minimum width=10mm]                       (hd)
  {Byte hard decision $\hat{\boldsymbol{\mu}}$ based on recevied signal};
\node[block, minimum width=10mm, below=2mm of hd]      (lm)
  {Semantic-prior model:\ $\Ps(\mu_i\mid\mathrm{ctx},\hat{\boldsymbol{\mu}})$};
\node[block, minimum width=10mm, below=2mm of lm]      (fu)
  { Bit-level and Byte-level score fusion};
\node[block, minimum width=30mm, below=4mm of fu, xshift=-20mm] (tb)
  {Bit-level TEPs $\mathcal{T}_b$};
\node[block, minimum width=30mm, below=4mm of fu, xshift= 20mm] (tB)
  {Byte-level TEPs $\mathcal{T}_b$};
\node[block, minimum width=60, below=4mm of tb, xshift=20mm]  (re)
  {Re-encoding TEPs and output optimal codeword};

\draw[->] (hd) -- (lm);
\draw[->] (lm) -- (fu);

\coordinate (br) at ($(fu.south)+(0,-2mm)$);
\draw            (fu)                 -- (br);
\draw            (tb.north |- br)     -- (tB.north |- br);
\draw[->]        (tb.north |- br)     -- (tb.north);
\draw[->]        (tB.north |- br)     -- (tB.north);

\coordinate (mg) at ($(re.north)+(0,2mm)$);
\draw            (tb.south)           -- (tb.south |- mg);
\draw            (tB.south)           -- (tB.south |- mg);
\draw            (tb.south |- mg)     -- (tB.south |- mg);
\draw[->]        (mg)                 -- (re.north);
\end{tikzpicture}
\caption{Decoding flow of \semOSD}
\label{fig:semosd-flow}

\vspace{-1.75em}
\end{figure}
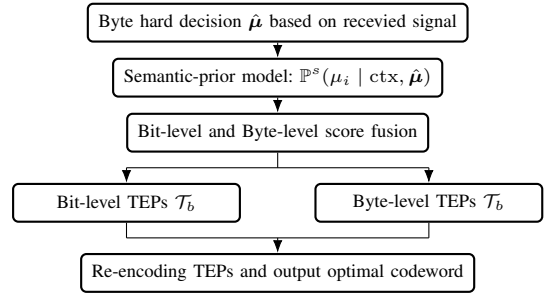

The flow of \semOSD\ is illustrated in Fig.~\ref{fig:semosd-flow}. The receiver forms the byte hard decision $\hat{\boldsymbol{\mu}}$, queries the semantic-prior model on $(\mathrm{ctx},\hat{\boldsymbol{\mu}})$, and fuses the channel posterior with the semantic prior into bit- and byte-level scores. The fused bit score forms the MRB and drives the bit-level TEP set $\Tbit$; the fused byte score drives the byte-level TEP set $\Tbyte$. All TEPs are re-encoded into candidate codewords, and the optimal one is output. This section introduces the semantic-channel fusion.


\vspace{-0.3em}
\subsection{Channel Posterior}\label{sec:semosd:channel-post}
\vspace{-0.3em}
The bit-level log-posterior under the observation model~\eqref{eq:obs} follows from the channel LLR $L_\ell = 2y_\ell/\hat\sigma^2$ as
\begin{align}\label{eq:bit-Pch}
\lc_\ell(\beta) = \log\Prob(c_\ell = \beta \mid y_\ell)= -\log\bigl(1 + e^{-(1-2\beta)\,L_\ell}\bigr),
\end{align}
for $\beta \in \Ftwo$. {Let $\mathbf{y}_i \triangleq (y_{8i}, \ldots, y_{8i+7})$ denote the eight channel observations of the $i$-th information byte $\mu_i$.} Treating the eight bits of $\mu_i$ as conditionally independent given $\mathbf{y}_i$, the byte-level  (character-level) log-posterior over $\Feight$ at byte position $i$ is
\begin{align}\label{eq:Pch}
\Lc_i(\nu) \triangleq \log\Pc(\mu_i \!\!=\!\! \nu \!\mid\! \mathbf{y}_i) =\! \sum_{j=0}^{7} \!\log\!\Prob\bigl(c_{8i+j} \!=\! \mathrm{bit}_j(\nu) \!\bigm|\! y_{8i+j}\bigr),
\end{align}
for $i \in \{0, \ldots, k-1\}$ and $\nu \in \Feight$, where $\mathrm{bit}_j(\nu)$ is the $j$-th bit of $\nu$. The conditional independence is exact on AWGN. On Gilbert--Elliott, it is approximate, since the bad-state indicator $S^{(g)}_\ell$ correlates adjacent bit observations; we accept this approximation under state-unaware reception. 

The channel hard decision for byte $\mu_i$ is $\hat\mu_i = \arg\max_\nu \Lc_i(\nu)$. We collect these into the information-byte hard-decision vector $\hat{\boldsymbol{\mu}} \triangleq (\hat\mu_0, \ldots, \hat\mu_{k-1}) \in \Feight^k$.

\vspace{-0.3em}
\subsection{Semantic Prior Model}\label{sec:semosd:semantic-prior}
\vspace{-0.3em}

\subsubsection{Architecture}\label{sec:semosd:semantic-prior:arch}
{The clean prefix $\mathrm{ctx}$ delivered by the system model of Section~\ref{sec:prelim:system} carries linguistic predictability about the current block. We exploit this through the \emph{semantic-prior model}, which consists of} a byte-level Transformer encoder followed by a per-position byte classifier. Semantic-prior model operates on $\mathrm{ctx}$ and the channel hard decision $\hat{\boldsymbol{\mu}}$ from~\eqref{eq:Pch}, with no access to the channel soft observations $\mathbf{y}$.

\begin{figure}[t]
\centering
\includegraphics[width=0.8\columnwidth]{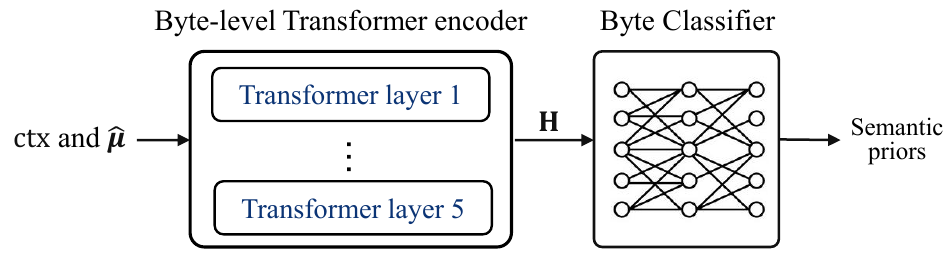}
\caption{Architecture of semantic prior model
}
\vspace{-1.25em}
\label{fig:SPM}
\end{figure}

The encoder is initialised from ByT5-small~\cite{xue2022byt5}, a byte-level Transformer of hidden dimension $d=1472$ pretrained on Common Crawl. At decoding time, the concatenation $(\mathrm{ctx},\hat{\boldsymbol{\mu}})\in\Feight^{L}$, where $L=L_c+k$ and $L_c$ is the byte length of $\mathrm{ctx}$, is mapped to per-position hidden states
\begin{equation}\label{eq:lm-enc}
\mathbf{H}=f_{\mathrm{Enc}}(\mathrm{ctx},\hat{\boldsymbol{\mu}};\theta)\in\mathbb{R}^{L\times d},
\end{equation}
under bidirectional self-attention. Let $ \mathbf{h}_i\in\mathbb{R}^{d}$ be the $i$-th row of $\mathbf{H}$. A linear layer $\mathbb{R}^{d}\to\mathbb{R}^{256}$ followed by softmax is then applied to the $k$ block-position hidden states $\mathbf{h}_{L_c+i}$, $i\in\{0,\ldots,k-1\}$, yielding at every information byte position $i$ a distribution $\Ps(\mu_i = \nu\mid\mathrm{ctx},\hat{\boldsymbol{\mu}})$ over $\nu\in\Feight$, which we call the \emph{semantic prior} at position $i$. The $L_c$ prefix-position hidden states are discarded; their role is exhausted by self-attention. The architecture is illustrated in Fig.~\ref{fig:SPM}.

The semantic prior is defined only at information byte positions $i\in\{0,\ldots,k-1\}$, equivalently at information bit positions $\ell\in\{0,\ldots,k_b-1\}$. We work with the log form
\begin{equation}\label{eq:Pby}
\Ls_i(\nu)\;\triangleq\;\log\Ps(\mu_i = \nu \mid \mathrm{ctx},\hat{\boldsymbol{\mu}}),\quad  \nu\in\Feight.
\end{equation}

An example of finding semantic priors is given in Fig.~\ref{fig:semantic-prior-example}.


\begin{figure}[t]
\centering
\footnotesize
\setlength{\tabcolsep}{4pt}
\begin{tabular}{@{}c c c l@{}}
\toprule
$i$ & $\hat\mu_i$ & $\nu^\star_i$
    & Top-$3$ entries of $\Ps(\mu_i = \nu\mid\mathrm{ctx},\hat{\boldsymbol{\mu}})$ \\
\midrule
$0$ & \texttt{0xE8} (\texttt{?})  & \texttt{h}
    & \textbf{\texttt{h}}\,$0.96$,\ \texttt{a}\,$0.02$,\ \texttt{o}\,$0.01$ \\
$1$ & \texttt{0x65} (\texttt{e})  & \texttt{e}
    & \textbf{\texttt{e}}\,$0.99$,\ \texttt{a}\,$0.005$,\ \texttt{i}\,$0.002$ \\
$2$ & \texttt{0x20} (\texttt{\textvisiblespace}) & \texttt{\textvisiblespace}
    & \textbf{\texttt{\textvisiblespace}}\,$0.99$,\ \texttt{,}\,$0.004$,\ \texttt{.}\,$0.002$ \\
$3$ & \texttt{0x73} (\texttt{s})  & \texttt{s}
    & \textbf{\texttt{s}}\,$0.98$,\ \texttt{c}\,$0.010$,\ \texttt{p}\,$0.004$ \\
$4$ & \texttt{0xEF} (\texttt{?})  & \texttt{o}
    & \textbf{\texttt{o}}\,$0.95$,\ \texttt{i}\,$0.02$,\ \texttt{a}\,$0.02$ \\
$5$ & \texttt{0x66} (\texttt{f})  & \texttt{f}
    & \textbf{\texttt{f}}\,$0.98$,\ \texttt{b}\,$0.008$,\ \texttt{p}\,$0.005$ \\
$6$ & \texttt{0x61} (\texttt{a})  & \texttt{a}
    & \textbf{\texttt{a}}\,$0.98$,\ \texttt{o}\,$0.010$,\ \texttt{e}\,$0.004$ \\
$7$ & \texttt{0x21} (\texttt{!})  & \texttt{\textvisiblespace}
    & \textbf{\texttt{\textvisiblespace}}\,$0.85$,\ \texttt{.}\,$0.08$,\ \texttt{,}\,$0.04$ \\
\bottomrule
\end{tabular}
\caption{Per-position semantic prior
$\Ps(\mu_i\mid\mathrm{ctx},\hat{\boldsymbol{\mu}})$ on a sentence ``The cat is sleeping on the sofa\,$\dots$''. The receiver
has clean prefix $\mathrm{ctx}=$``The cat is sleeping on t'' and observes the noisy hard decision $\hat{\boldsymbol{\mu}} = \text{``?e\ s?fa!''}$. Here, $\nu_i^*$ represents the top-1 byte.}
\label{fig:semantic-prior-example}
\vspace{-1em}
\end{figure}

\subsubsection{Training}\label{sec:semosd:semantic-prior:train}

We fine-tune the classifier and the upper encoder blocks on the SNLI hypotheses subset~\cite{bowman2015large}. Each training sample takes an English sentence of $kG$ bytes, partitions it into $G$ groups of $k$ bytes, picks one group $g\in\{1,\ldots,G\!-\!1\}$ uniformly, and corrupts it bit-wise at rate $p=0.1$ to obtain a noisy version $\tilde{\boldsymbol{\mu}}^{(g)}\in\Feight^k$. The model is queried with $\mathrm{ctx}^{(g)}$ as the prefix and $\tilde{\boldsymbol{\mu}}^{(g)}$ in place of the channel hard decision; the per-sample loss is the per-position cross-entropy
\begin{equation}\label{eq:lm-loss}
\mathcal{L} = -\sum_{i=0}^{k-1}\log\Ps\bigl(\mu_i = \mu^{(g)}_i \,\bigm|\, \mathrm{ctx}^{(g)},\,\tilde{\boldsymbol{\mu}}^{(g)}\bigr).
\end{equation}
The procedure uses no information about the channel code, modulation, or decoder, so the resulting model is a code-agnostic byte denoiser.

\vspace{-0.3em}
\subsection{Fused Bit-Level and Byte-Level Score}\label{sec:semosd:fusion}
\vspace{-0.3em}

{\semOSD\ performs decoding at the bit level, so the byte-level distribution \eqref{eq:Pby} must be reduced to per-bit form. Within $n_b$ codeword bits, let $\ell = 8i+j$ for information byte $i$ and bit $j\in\{0,\ldots,7\}$ within the byte. The bit-level semantic prior at an information bit position $\ell\in\{0,\ldots,k_b-1\}$ is obtained by summing the byte distribution over byte values whose $j$-th bit equals $\beta$; that is,
\begin{equation}\label{eq:bitmarg}
\ls_\ell(\beta) \;\triangleq\; \log\!\!\sum_{\substack{\nu\in\Feight\\\mathrm{bit}_j(\nu)=\beta}}\!\!\Ps(\mu_i = \nu \mid \mathrm{ctx},\hat{\boldsymbol{\mu}}),\quad \beta\in\Ftwo.
\end{equation}}

The quantity $\lc_\ell$ from \eqref{eq:bit-Pch} is a bit-level channel log-likelihood, while $\ls_\ell$ inherits the softmax magnitude of the LM classification head and can be much larger. We row-normalise each by its row maximum, i.e.,
\begin{equation}\label{eq:tilde_c}
\tlc_\ell(\beta)=\lc_\ell(\beta)-\max_{\beta'\in\Ftwo}\lc_\ell(\beta'),
\end{equation}
and
\begin{equation}\label{eq:tilde_s}
\tls_\ell(\beta)=\ls_\ell(\beta)-\max_{\beta'\in\Ftwo}\ls_\ell(\beta').
\end{equation}

The \emph{fused bit-level score} is a convex combination on information bits and the channel-only score on parity bits,
\begin{equation}\label{eq:fusion}
\lf_\ell(\beta)=
\begin{cases}
\alpha\,\tlc_\ell(\beta)+(1-\alpha)\,\tls_\ell(\beta), & \ell<k_b,\\[2pt]
\tlc_\ell(\beta), & \ell\ge k_b,
\end{cases}
\end{equation}
with mixing weight $\alpha\in[0,1]$. Setting $\alpha = 1$ recovers Fossorier OSD, and $\alpha = 0$ ranks information bits by the LM evidence alone. The mixing weight $\alpha$ is selected empirically per channel; see Section~\ref{sec:exp:setup}.

{The same fusion applies at the byte level. Let $\tLc_i$ and $\tLs_i$ be the row-normalised forms of $\Lc_i$ and $\Ls_i$ from~\eqref{eq:Pch}--\eqref{eq:Pby}, respectively. The fused byte score on information bytes is
\begin{equation}\label{eq:fused-byte}
L_i(\nu) = \alpha\,\tLc_i(\nu) + (1-\alpha)\,\tLs_i(\nu), \quad  \nu\in\Feight,
\end{equation}
for $i\in\{0,\ldots,k-1\}$, with the same mixing weight $\alpha$.}

\section{The \semOSD\ Algorithm}\label{sec:semosd}

\vspace{-0.3em}

\subsection{Most-Reliable Basis}\label{sec:semosd:mrb}
\vspace{-0.3em}

{Define the fused bit-level LLR
\begin{equation}\label{eq:fusedllr}
\Lambda_\ell \;\triangleq\; \lf_\ell(0) - \lf_\ell(1), \quad \ell\in\{0,\ldots,n_b-1\},
\end{equation}
and the bit reliability $r_\ell = |\Lambda_\ell|$.} {Let $\pi$ be the permutation of $\{0,\ldots,n_b-1\}$ that sorts the bit positions by $r_\ell$ in decreasing order; that is, $r_{\pi(0)}\ge r_{\pi(1)}\ge\cdots\ge r_{\pi(n_b-1)}$.} We apply $\pi$ to the columns of $\mathbf{G}_b$ and reduce the resulting matrix to systematic form $\mathbf{G}_s = [\mathbf{I}_{k_b} \mid \mathbf{P}']$ over $\Ftwo$; the leading $k_b$ permuted positions then form the MRB. If the leading $k_b$ submatrix is rank-deficient, OSD restores full rank by swapping the offending column with the next-most-reliable position outside the leading block~\cite{fossorier1995soft}. RS$(16,8)$ is maximum-distance-separable, so no swap is ever needed; in contrast, BCH$(127,64)$ produces occasional swaps that affect only the column ordering.

{Let $\mathbf{u}_0\in\Ftwo^{k_b}$ be the hard decision under $\Lambda_\ell$ at the MRB,
\begin{equation}\label{eq:u0}
u_{0,\ell} \;=\; \begin{cases} 0, & \Lambda_{\pi(\ell)} \ge 0,\\ 1, & \Lambda_{\pi(\ell)} < 0, \end{cases} \quad \ell\in\{0,\ldots,k_b-1\}.
\end{equation}
The TEP evaluation process of conventional OSD can enumerate bit perturbations of $\mathbf{u}_0$.} {The fused reliability $r_\ell$ thus reflects both channel reliability and semantic reliability.}

\vspace{-0.3em}
\subsection{Two TEP Families}\label{sec:semosd:tep}
\vspace{-0.3em}

Conventional order-$m$ OSD succeeds only when the MRB error has Hamming weight at most $m$. This range can be exceeded when the channel is poor or when burst errors concentrate flips inside a few bytes. \semOSD\ addresses this by employing two TEP families: a bit-flip family at order $m$ for low-weight MRB errors, and a byte-substitution family that reaches such byte-clustered errors.

\subsubsection{Bit-flip family $\Tbit$}

The bit-flip family operates over the MRB, which is standard Fossorier OSD search. A bit-flip TEP is a binary vector $\mathbf{e}\in\Ftwo^{k_b}$ whose support marks which MRB bits of $\mathbf{u}_0$ are hypothesised to be in error.

\begin{definition}[Bit-flip family]\label{def:Tbit}
For an integer $m\ge 0$, the bit-flip TEP family is
\begin{equation}\label{eq:Tbit}
\Tbit \;=\; \bigl\{\mathbf{e}\in\Ftwo^{k_b} \,:\, \mathrm{wt}(\mathbf{e})\le m\bigr\},
\end{equation}
where $\mathrm{wt}(\mathbf{e})$ is the Hamming weight.
\end{definition}

\subsubsection{Byte-substitution family $\Tbyte$}
The byte-substitution family operates over the original information bytes, not the MRB. It is parametrised by two integers, $\omega\ge 1$ and $T\ge 1$.

For each information position $i\in\{0,\ldots,k-1\}$, \semOSD sorts the $256$ byte values by the fused byte score $L_i(\nu)$ of~\eqref{eq:fused-byte} in decreasing order. Let $\nu^\star_i\in\Feight$ denote the top-ranked value, and $\nu^{(1)}_i,\ldots,\nu^{(T)}_i\in\Feight$ the next $T$ alternatives. We denote $\boldsymbol{\mu}^\star=(\nu^\star_0,\ldots,\nu^\star_{k-1})$ for the top-ranked information word.

\begin{definition}[Byte-substitution family]\label{def:Tbyte}
A byte-substitution TEP is specified by two steps:
\begin{enumerate}
\item pick a subset $\mathcal{S}\subseteq\{0,\ldots,k-1\}$ of up to $\omega$ byte positions;
\item at each $i\in\mathcal{S}$, pick a replacement value $v_i\in\{\nu^{(1)}_i,\ldots,\nu^{(T)}_i\}$. We denote $\mathbf{v}=(v_i)_{i\in\mathcal{S}}$.
\end{enumerate}
The choice $(\mathcal{S},\mathbf{v})$ defines a byte-level TEP $\boldsymbol{\eta}_{\mathcal{S},\mathbf{v}}\in\Feight^k$ on the information bytes,
\begin{equation*}
(\boldsymbol{\eta}_{\mathcal{S},\mathbf{v}})_i \;=\; \begin{cases} v_i\oplus\nu^\star_i, & i\in\mathcal{S},\\ 0, & i\notin\mathcal{S}, \end{cases}
\end{equation*}
and a corresponding candidate information word $\boldsymbol{\mu}^\star\oplus\boldsymbol{\eta}_{\mathcal{S},\mathbf{v}}$.  Family $\Tbyte$ collects all such TEPs over valid choices of $(\mathcal{S},\mathbf{v})$, including the all-zero TEP at $\mathcal{S}=\emptyset$.
\end{definition}

A TEP $\boldsymbol{\eta}\in \Tbyte$ has byte-weight up to $\omega$, corresponding to up to $8\omega$ bit flips when expanded.

\vspace{-0.3em}
\subsection{Re-encoding and Output}\label{sec:semosd:reenc}
\vspace{-0.3em}

The TEPs in $\Tbit$ and $\Tbyte$ live in different spaces, i.e. bit-level at the MRB versus byte-level at the information bytes, but they produce candidate codewords in the same $\Ftwo^{n_b}$ space.

For a bit-flip TEP $\mathbf{e}\in\Tbit$, the candidate codeword is
\begin{equation}\label{eq:reenc-bit}
\mathbf{x}(\mathbf{e}) \;=\; \pi^{-1}\bigl((\mathbf{u}_0 \oplus \mathbf{e})\mathbf{G}_s\bigr),
\end{equation}
where $(\mathbf{u}_0 \oplus \mathbf{e})\mathbf{G}_s$ is a permuted codeword and $\pi^{-1}$ converts it to the original codeword basis. For a byte-substitution TEP $\boldsymbol{\eta}\in\Tbyte$, the candidate information word $\boldsymbol{\mu}^\star\oplus\boldsymbol{\eta}\in\Feight^k$ has bit expansion $\mathbf{u}\in\Ftwo^{k_b}$ given by $u_{8i+j}=\mathrm{bit}_j((\boldsymbol{\mu}^\star\oplus\boldsymbol{\eta})_i)$, and the candidate codeword in the original basis is
\begin{equation}\label{eq:reenc-byte}
\mathbf{x}(\boldsymbol{\eta}) \;=\; \mathbf{u}\,\mathbf{G}_b.
\end{equation}

Each candidate $\mathbf{x}\in \mathcal{X}= \bigl\{\mathbf{x}(\mathbf{e}):\mathbf{e}\in\Tbit\bigr\}\cup\bigl\{\mathbf{x}(\boldsymbol{\eta}):\boldsymbol{\eta}\in\Tbyte\bigr\}$ is scored against the fused bit-level score,
\begin{equation}\label{eq:dist}
d(\mathbf{x}) \;=\; -\sum_{\ell=0}^{n_b-1}\lf_\ell(x_\ell).
\end{equation}
The score uses channel evidence on all $n_b$ codeword bits and LM evidence on the $k_b$ information bits. The decoding output is the codeword which minimizes $d(\mathbf{x})$, i.e., $\hat{\mathbf{c}} \;=\; \arg\min_{\mathbf{x} \in \mathcal{X}}\,d(\mathbf{x}).$

The algorithm of Sem-OSD is summarized in Algorithm \ref{alg:semosd}.

\begin{algorithm}[h]
\small
\caption{Semantic Ordered Statistics Decoding}
\label{alg:semosd}
\begin{algorithmic}[1]
\REQUIRE Received $\mathbf{y}\in\mathbb{R}^{n_b}$, block-average variance $\hat\sigma^2$, prefix $\mathrm{ctx}$, generator $\mathbf{G}_b$, parameters $(\alpha,m,\omega,T)$
\ENSURE Decoded codeword $\hat{\mathbf{c}}\in\Ftwo^{n_b}$.
\STATE Compute the byte-level posteriors $\Lc_i(\nu)$ via~\eqref{eq:Pch} and $\Ls_i(\nu)$ via~\eqref{eq:Pby} for $i\in\{0,\ldots,k-1\}$ (one LM forward pass)
\STATE Compute the fused bit-level score $\lf_\ell(\beta)$ via~\eqref{eq:fusion} and the fused byte score $L_i(\nu)$ via~\eqref{eq:fused-byte}
\STATE Set $\Lambda_\ell\leftarrow\lf_\ell(0)-\lf_\ell(1)$ 
\STATE Obtain permutation $\pi$ by sorting $|\Lambda_\ell|$ in decreasing order.
\STATE Construct systematic $\mathbf{G}_s$ by permuting $\mathbf{G}_b$
\STATE Set $u_{0,\ell}\leftarrow\mathbf{1}[\Lambda_{\pi(\ell)}<0]$ for $\ell=0,\ldots,k_b\!-\!1$.
\STATE For each information position $i\in\{0,\ldots,k-1\}$, sort $\Feight$ by $L_i(\cdot)$ and extract the top value $\nu^\star_i$ and the next $T$ alternatives $\nu^{(1)}_i,\ldots,\nu^{(T)}_i$; assemble $\boldsymbol{\mu}^\star=(\nu^\star_0,\ldots,\nu^\star_{k-1})$.
\STATE Construct $\Tbit$ via~\eqref{eq:Tbit} and $\Tbyte$ per Definition~\ref{def:Tbyte}
\STATE Re-encode each TEP from  $\Tbit$ and $\Tbyte$ via~\eqref{eq:reenc-bit} and~\eqref{eq:reenc-byte}
\STATE \textbf{return} $\hat{\mathbf{c}}$ which minimizes \eqref{eq:dist}
\end{algorithmic}
\end{algorithm}

\vspace{-0.3em}
\subsection{Complexity Overhead}\label{sec:semosd:complexity}
\vspace{-0.3em}

Each \semOSD decoding can evaluate up to
\begin{equation}\label{eq:Ntep}
\Ntep = |\Tbit|+|\Tbyte| = \sum_{w=0}^{m}\binom{k_b}{w} + \sum_{w=0}^{\omega}\binom{k}{w}T^w
\end{equation}
candidate codewords. {For binary BCH$(127,64)$ at $(m,\omega,T)=(4,2,16)$, this gives $|\Tbit|=679{,}121$, $|\Tbyte|=7{,}297$, and $\Ntep=686{,}418$, adding $7{,}297$ TEPs over Fossorier OSD's $679{,}121$ at order $m=4$.} 

The complexity of \semOSD roughly decomposes as
\begin{align}\label{eq:runtime}
C_{\mathrm{Sem-OSD}} = C_{\mathrm{LM}} + C_{\mathrm{Byte}} +  C_{\mathrm{OSD}}(\Ntep),
\end{align}
where $C_{\mathrm{LM}}$ is complexity of one LM forward pass, $C_{\mathrm{Byte}} = \mathcal{O}(256k)$ denotes the byte scoring overhead, and $C_{\mathrm{OSD}}(\Ntep)$ is the complexity of standard OSD process that evaluates $\Ntep$ TEPs.
Therefore \semOSD\ is more expensive than conventional OSD due to the LM forward pass and the evaluation of byte TEPs.

The byte family is empirically needed only on burst channels. Section~\ref{sec:exp:awgn} shows that on AWGN the bit-flip family alone tracks the full \semOSD\ curve closely. A practical deployment may therefore disable $\Tbyte$ on AWGN channels, and the complexity becomes
\begin{equation}\label{eq:runtime:AWGN}
C_{\mathrm{bit-TEP-only}} = C_{\mathrm{LM}} + C_{\mathrm{OSD}}(|\Tbit|),
\end{equation}
i.e., conventional OSD plus a single LM forward pass.  The cost $C_{\mathrm{LM}}$ highly depends on the GPU and implementation. 

A strategy to keep the LM cost minimal is that the receiver first runs the code's native decoder on the received signal and invokes \semOSD only when that decoder fails. Suitable choices include BM~\cite{berlekamp1968algebraic} for RS and BCH, or BP for LDPC.

\vspace{-0.5em}
\section{Experiments}\label{sec:exp}
\vspace{-0.5em}
\subsection{Setup}\label{sec:exp:setup}
\vspace{-0.3em}
{We instantiate \semOSD\ on shortened RS$(16,8)$ over $\Feight$ and on binary BCH$(127,64)$. Both codes are evaluated on AWGN and on a Gilbert--Elliott channel with $\pi_B=0.10$, burst length $\bar L=16$ bits, and $\rho^2=100$.}

{The source corpus is the SNLI hypotheses subset~\cite{bowman2015large}, filtered to $60$--$64$-character sentences. For each test sentence, we sample $g$ uniformly from $\{2,\ldots,8\}$ so that decoding sees a random prefix of $g-1$ clean $8$-byte segments. The semantic-prior model is fine-tuned for five epochs at learning rate $10^{-4}$, batch size $32$, with the upper ByT5 encoder transformer blocks and classification head trained jointly and the bottom four transformer blocks frozen.}

{\semOSD\ uses $(\omega,T)=(2,16)$ throughout, with $m=2$ on Gilbert--Elliott RS$(16,8)$, $m=3$ on AWGN RS$(16,8)$, and $m=4$ on AWGN BCH$(127,64)$. The mixing weight $\alpha$ is selected as $\alpha=0.1$ on Gilbert--Elliott and $\alpha=0.5$ on AWGN{, identified empirically as the best-performing settings}. Baselines are BM and Fossorier OSD at the same $m$.}

\vspace{-0.3em}
\subsection{Error Rate Performance}\label{sec:exp:awgn}
\vspace{-0.3em}

\subsubsection{AWGN Channel}
\begin{figure}[t]
\centering
\begin{tikzpicture}
\begin{axis}[
    width=0.96\linewidth, height=5.5cm,
    xlabel={$E_b/N_0$ (dB)}, ylabel={BLER},
    ymode=log, log basis y=10,
    ymin=1e-6, ymax=1.0, xmin=-0.2, xmax=3.2,
    tick label style={font=\footnotesize},
    xmajorgrids,
    ymajorgrids,
    yminorgrids,
    major grid style={dotted,black},
    minor grid style={dotted},
    legend style={at={(0.025,0.025)}, anchor=south west, legend cell align=left, align=left, draw=white!15!black,font = \scriptsize	,row sep=-1pt, legend columns=1, , fill opacity=0.5, text opacity=1}
]
\addplot[mark=*, mark size=1.5pt, color=black] coordinates {
  (0.0, 1.0) (0.5, 1.0) (1.0, 1.0) (1.5, 1.0) (2.0, 0.9901) (2.5, 0.9804) (3.0, 0.9804)
};
\addlegendentry{BM~\cite{berlekamp1968algebraic}}
\addplot[mark=square*, mark size=1.5pt, color=blue!75!black] coordinates {
  (0.0, 0.4902) (0.5, 0.2985) (1.0, 0.1176) (1.5, 0.0535) (2.0, 0.0126) (2.5, 0.0025) (3.0, 0.0004)
};
\addlegendentry{OSD ($m=3$)~\cite{fossorier1995soft}}
\addplot[mark=triangle*, mark size=2.0pt, color=violet!80!black, dashed] coordinates {
  (0.0, 0.50) (0.5, 0.40) (1.0, 0.32) (1.5, 0.25) (2.0, 0.19) (2.5, 0.145) (3.0, 0.108)
};
\addlegendentry{\semOSD\ ($\Tbyte$ only)}
\addplot[mark=otimes*, mark size=2.0pt, color=teal!80!black, dashdotted] coordinates {
  (0.0, 0.0645) (0.5, 0.024) (1.0, 0.0099) (1.5, 0.003) (2.0, 0.0005) (2.5, 7e-5) (3.0, 9e-6)
};
\addlegendentry{\semOSD\ ($\Tbit$ only)}
\addplot[mark=diamond*, mark size=2.5pt, color=red] coordinates {
  (0.0, 0.0595) (0.5, 0.0231) (1.0, 0.0088) (1.5, 0.0025) (2.0, 0.0004) (2.5, 5e-5) (3.0, 5e-6)
};
\addlegendentry{\semOSD}

\addplot[mark=none, color=gray!70!black, dashed, very thick] coordinates {
  (0.00, 0.4266) (0.25, 0.3273) (0.50, 0.2375) (0.75, 0.1618) (1.00, 0.1027)
  (1.25, 0.0603) (1.50, 0.0324) (1.75, 0.0158) (2.00, 0.0069) (2.25, 0.0027)
  (2.50, 8.95e-4) (2.75, 2.57e-4) (3.00, 6.17e-5)
};
\addlegendentry{Normal Approximation Bound}

\end{axis}
\end{tikzpicture}
\vspace{-0.5em}
\caption{BLER performance on RS$(16,8)$ over AWGN. }
\vspace{-0.5em}
\label{fig:bler-awgn}
\end{figure}
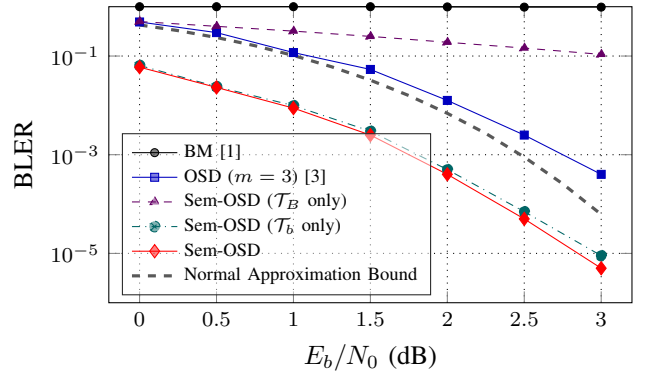

Figure~\ref{fig:bler-awgn} reports BLER on the AWGN channel for $(16,8)$ RS code, where \semOSD\ has parameters $(m,\omega,T,\alpha)=(3,2,16,0.5)$. As can be seen, {\semOSD\ significantly reduces BLER over Fossorier OSD by approximately $20\times$ across $0$--$3$\,dB. The bit-flip family $\Tbit$ alone already captures most of the AWGN error pattern, while the $\Tbyte$-only variation plateaus near $0.1$, since AWGN errors are scattered across many bytes rather than clustered inside a few. The byte-substitution family is therefore mismatched to the AWGN error structure.}

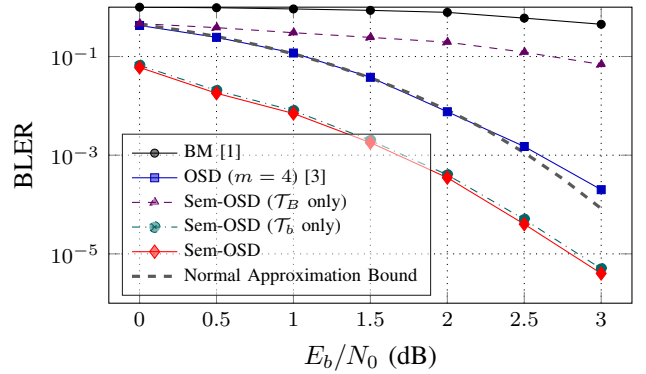
\begin{figure}[t]
\centering
\begin{tikzpicture}
\begin{axis}[
    width=0.96\linewidth, height=5.5cm,
    xlabel={$E_b/N_0$ (dB)}, ylabel={BLER},
    ymode=log, log basis y=10,
    ymin=1e-6, ymax=1.0, xmin=-0.2, xmax=3.2,
    tick label style={font=\footnotesize},
    xmajorgrids,
    ymajorgrids,
    yminorgrids,
    major grid style={dotted,black},
    minor grid style={dotted},
    legend style={at={(0.025,0.025)}, anchor=south west, legend cell align=left, align=left, draw=white!15!black,font = \scriptsize	,row sep=-1pt, legend columns=1 , fill opacity=0.5, text opacity=1}
]
\addplot[mark=*, mark size=1.5pt, color=black] coordinates {
  (0.0, 1.0000) (0.5, 0.9740) (1.0, 0.9202) (1.5, 0.8621) (2.0, 0.7853) (2.5, 0.6000) (3.0, 0.4505)
};
\addlegendentry{BM~\cite{berlekamp1968algebraic}}
\addplot[mark=square*, mark size=1.5pt, color=blue!75!black] coordinates {
  (0.0, 0.4261) (0.5, 0.2447) (1.0, 0.1178) (1.5, 0.0379) (2.0, 0.0076) (2.5, 0.0015) (3.0, 0.0002)
};
\addlegendentry{OSD ($m=4$)~\cite{fossorier1995soft}}
\addplot[mark=triangle*, mark size=2.0pt, color=violet!80!black, dashed] coordinates {
  (0.0, 0.4644) (0.5, 0.3846) (1.0, 0.3036) (1.5, 0.2439) (2.0, 0.1951) (2.5, 0.1222) (3.0, 0.0691)
};
\addlegendentry{\semOSD\ ($\Tbyte$ only)}

\addplot[mark=otimes*, mark size=2.0pt, color=teal!80!black, dashdotted] coordinates {
  (0.0, 0.0662) (0.5, 0.0205) (1.0, 0.008) (1.5, 0.002) (2.0, 4e-4) (2.5, 5e-5) (3.0, 5e-6)
};
\addlegendentry{\semOSD\ ($\Tbit$ only)}
\addplot[mark=diamond*, mark size=2.5pt, color=red] coordinates {
  (0.0, 0.060) (0.5, 0.018) (1.0, 0.007) (1.5, 0.0018) (2.0, 3.5e-4) (2.5, 4e-5) (3.0, 4e-6)
};
\addlegendentry{\semOSD}

\addplot[mark=none, color=gray!70!black, dashed, very thick] coordinates {
  (0.00, 0.4474) (0.25, 0.3470) (0.50, 0.2548) (0.75, 0.1759) (1.00, 0.1134)
  (1.25, 0.0677) (1.50, 0.0370) (1.75, 0.0184) (2.00, 0.0082) (2.25, 0.0033)
  (2.50, 0.001125834145464) (2.75, 3.335016628889930e-04) (3.00, 8.293647019708146e-05)
};
\addlegendentry{Normal Approximation Bound}

\end{axis}
\end{tikzpicture}
\vspace{-0.5em}
\caption{BLER Performance on binary BCH$(127,64)$ over AWGN.}
\vspace{-1em}
\label{fig:bler-awgn-bch}
\end{figure}

{Figure~\ref{fig:bler-awgn-bch} repeats the AWGN experiment on BCH$(127,64)$ with $m=4$. \semOSD\ has parameters $(m,\omega,T,\alpha)=(4,2,16,0.5)$. As seen, \semOSD improves BLER over Fossorier OSD by $7\times$ at $0$\,dB and by $22\times$ at $2$\,dB. Both $\Tbit$-only and $\Tbyte$-only variations behave similarly to RS.}

{Notably, the BLER curves of \semOSD\ on both RS$(16,8)$ and BCH$(127,64)$ fall below the normal-approximation finite-blocklength bound~\cite{PPV2010l} at the corresponding code dimensions, which is derived under a uniform-source assumption. \semOSD\ exceeds the bound by exploiting the source's non-uniform distribution.}

\subsubsection{Gilbert--Elliott Channel}\label{sec:exp:ge}

\begin{figure}[t]
\centering
\begin{tikzpicture}
\begin{axis}[
    width=0.96\linewidth, height=5.5cm,
    xlabel={$E_b/N_0$ (dB)}, ylabel={BLER},
    ymode=log, log basis y=10,
    ymin=1e-6, ymax=1.0, xmin=3.5, xmax=14.5,
    tick label style={font=\footnotesize},
    xmajorgrids,
    ymajorgrids,
    yminorgrids,
    major grid style={dotted,black},
    minor grid style={dotted},
    legend style={at={(0.025,0.025)}, anchor=south west, legend cell align=left, align=left, draw=white!15!black,font = \scriptsize	,row sep=-1pt, legend columns=1 , fill opacity=0.5, text opacity=1}
]
\addplot[mark=*, mark size=1.5pt, color=black] coordinates {
  (4.0, 0.122) (6.0, 0.096) (8.0, 0.081) (10.0, 0.059) (12.0, 0.025) (14.0, 0.0046)
};
\addlegendentry{BM~\cite{berlekamp1968algebraic}}
\addplot[mark=square*, mark size=1.5pt, color=blue!75!black] coordinates {
  (4.0, 0.114) (6.0, 0.102) (8.0, 0.044) (10.0, 0.0093) (12.0, 8.5e-4) (14.0, 6e-5)
};
\addlegendentry{OSD ($m=2$) ~\cite{fossorier1995soft}}
\addplot[mark=triangle*, mark size=2.0pt, color=violet!80!black, dashed] coordinates {
  (4.0, 0.046) (6.0, 0.032) (8.0, 0.021) (10.0, 0.0095) (12.0, 0.0015) (14.0, 2e-4)
};
\addlegendentry{\semOSD\ ($\Tbyte$ only)}
\addplot[mark=otimes*, mark size=2.0pt, color=teal!80!black, dashdotted] coordinates {
  (4.0, 0.092) (6.0, 0.054) (8.0, 0.026) (10.0, 0.0083) (12.0, 5e-4) (14.0, 3e-5)
};
\addlegendentry{\semOSD\ ($\Tbit$ only)}
\addplot[mark=diamond*, mark size=2.5pt, color=red] coordinates {
  (4.0, 0.027) (6.0, 0.015) (8.0, 0.0049) (10.0, 0.0012) (12.0, 1.5e-4) (14.0, 5e-6)
};
\addlegendentry{\semOSD\ }
\end{axis}
\end{tikzpicture}
\vspace{-0.5em}
\caption{BLER on RS$(16,8)$ over Gilbert--Elliott burst-error channel.}
\label{fig:bler-ge}
\end{figure}
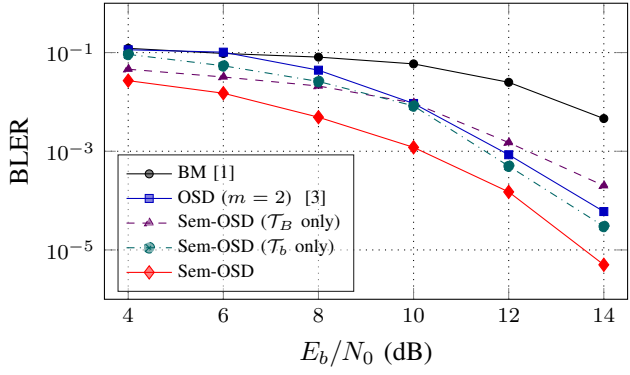

Figure~\ref{fig:bler-ge} reports BLER versus $E_b/N_0$ on RS$(16,8)$ over the Gilbert--Elliott channel.
{\semOSD\ outperforms BM and OSD over the entire SNR range, reducing BLER over OSD by $4.5\times$ at $4$\,dB and by one order of magnitude at $12$\,dB. The advantage comes from both TEP families being effective on burst errors. Byte-clustered errors are repaired by $\Tbyte$, while the residual scattered errors fall within $\Tbit$'s reach.} { $\Tbyte$-only dominates at low SNR ($4$--$8$\,dB), where bursts produce byte-clustered errors that a single byte substitution repairs. $\Tbit$-only dominates at high SNR ($\ge 10$\,dB), where bursts are rare and residual errors are isolated bit flips.}

\subsubsection{Semantic Score}

Table~\ref{tab:sbert} reports the SBERT cosine similarity $s\in[-1,1]$~\cite{reimers2019sentence} between the decoded text and the source, averaged over the test split. Higher is better; $s\!\to\!1$ corresponds to near-source semantic recovery.

Across all  settings, \semOSD\ achieves the highest SBERT score at every $E_b/N_0$ tested. {At $E_b/N_0=4$\,dB on GE, \semOSD\ retains an SBERT similarity of $s=0.980$ despite a non-trivial $\mathrm{BLER}=2.74\!\times\!10^{-2}$, indicating that some residual block errors land near the source in the semantic embedding space rather than at arbitrary byte vectors. The LM-driven byte-level TEPs can biases \semOSD\ toward semantically plausible codewords. The same trend holds on BCH$(127,64)$. }

\begin{table}[t]
\footnotesize
\caption{SBERT cosine similarity across the different experiments. Values close to $1$ indicate near-source semantic recovery.}
\label{tab:sbert}
\centering\small
\setlength{\tabcolsep}{4pt}
\begin{tabular}{c|ccccc}
\hline
$E_b/N_0$ & BM & OSD & $\Tbyte$-only & $\Tbit$-only & \semOSD \\
\hline
\multicolumn{6}{c}{\emph{RS$(16,8)$ over AWGN, $m=3$}} \\
\hline
$1$\,dB & $0.275$ & $0.907$ & $0.756$ & $0.992$ & $\mathbf{0.993}$ \\
$2$\,dB & $0.300$ & $0.990$ & $0.858$ & $1.000$ & $\mathbf{1.000}$ \\
$3$\,dB & $0.353$ & $1.000$ & $0.925$ & $1.000$ & $\mathbf{1.000}$ \\
\hline
\multicolumn{6}{c}{\emph{BCH$(127,64)$ over AWGN, $m=4$}} \\
\hline
$1$\,dB & $0.311$ & $0.906$ & $0.784$ & $0.994$ & $\mathbf{0.995}$ \\
$2$\,dB & $0.427$ & $0.994$ & $0.865$ & $1.000$ & $\mathbf{1.000}$ \\
$3$\,dB & $0.686$ & $1.000$ & $0.951$ & $1.000$ & $\mathbf{1.000}$ \\
\hline
\multicolumn{6}{c}{\emph{RS$(16,8)$ over Gilbert--Elliott, $m=2$}} \\
\hline
$4$\,dB  & $0.932$ & $0.908$ & $0.969$ & $0.931$ & $\mathbf{0.980}$ \\
$8$\,dB  & $0.955$ & $0.966$ & $0.986$ & $0.979$ & $\mathbf{0.996}$ \\
$12$\,dB & $0.986$ & $0.998$ & $0.999$ & $1.000$ & $\mathbf{1.000}$ \\
\hline
\end{tabular}
\vspace{-0.5em}
\end{table}

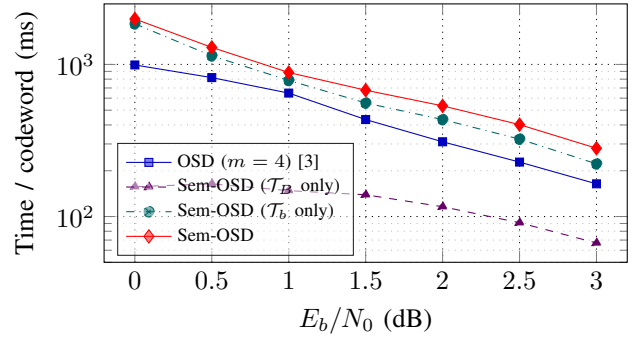
\begin{figure}[t]
\centering
\begin{tikzpicture}
\begin{axis}[
    width=0.96\linewidth, height=5cm,
    xlabel={$E_b/N_0$ (dB)}, ylabel={Time / codeword (ms)},
    ymode=log, log basis y=10,
    ymin=50, ymax=2500, xmin=-0.2, xmax=3.2,
    xmajorgrids,
    ymajorgrids,
    yminorgrids,
    major grid style={dotted,black},
    minor grid style={dotted},
    legend style={at={(0.025,0.025)}, anchor=south west, legend cell align=left, align=left, draw=white!15!black,font = \scriptsize	,row sep=-1pt, legend columns=1 , fill opacity=0.5, text opacity=1}
]
\addplot[mark=square*, mark size=1.5pt, color=blue!75!black] coordinates {
  (0.0, 992) (0.5, 819) (1.0, 648) (1.5, 434) (2.0, 310) (2.5, 228) (3.0, 164)
};
\addlegendentry{OSD ($m=4$) \cite{fossorier1995soft}}
\addplot[mark=triangle*, mark size=2.0pt, color=violet!80!black, dashed] coordinates {
  (0.0, 157) (0.5, 164) (1.0, 148) (1.5, 139) (2.0, 116) (2.5, 91) (3.0, 67)
};
\addlegendentry{\semOSD\ ($\Tbyte$ only)}
\addplot[mark=otimes*, mark size=2.0pt, color=teal!80!black, dashdotted] coordinates {
  (0.0, 1850) (0.5, 1139) (1.0, 785) (1.5, 558) (2.0, 434) (2.5, 323) (3.0, 222)
};
\addlegendentry{\semOSD\ ($\Tbit$ only)}
\addplot[mark=diamond*, mark size=2.5pt, color=red] coordinates {
  (0.0, 1990) (0.5, 1294) (1.0, 885) (1.5, 677) (2.0, 534) (2.5, 402) (3.0, 281)
};
\addlegendentry{\semOSD}
\end{axis}
\end{tikzpicture}
\vspace{-0.5em}
\caption{Average decoding time per codeword on AWGN for BCH$(127,64)$ at order $m=4$.}
\label{fig:dectime}
\vspace{-1.5em}
\end{figure}

\vspace{-0.3em}
\subsection{Decoding Latency}
\vspace{-0.3em}

Figure~\ref{fig:dectime} shows the per-codeword decoding time on AWGN BCH$(127,64)$ at $m=4$ and \semOSD  $(4,2,16,0.5)$. Both the OSD baseline and \semOSD\ enumerate TEPs under the PB-OSD stopping rule~\cite{yue2021probability}, which terminates the search once a sufficiently likely candidate is found.  Full \semOSD\ takes $2000$ ms at $0$\,dB and $280$ ms at $3$\,dB. We note that \semOSD\ runs about $2\times$ slower than OSD, since it invokes the LM. Nevertheless, it trades that for the $1.5$\,dB coding gain over the normal-approximation bound, as demonstrated in Section~\ref{sec:exp:awgn}.

{If we run BM first and invoke \semOSD\ only when BM fails, the LM is invoked on only $30\%$ of blocks at $E_b/N_0=3$\,dB on AWGN and $1\%$ of blocks at $E_b/N_0=14$\,dB on Gilbert--Elliott channel. We omit the detailed results due to space limit.}

\vspace{-0.8em}
\section{Conclusion}\label{sec:conclusion}
\vspace{-0.3em}
{We introduced \semOSD, a soft decoder for short byte-aligned linear block codes that fuses a byte-level language-model (LM) prior with channel reliability. \semOSD\ enumerates two complementary test-error-pattern (TEP) families: a bit-flip family at the most-reliable basis (MRB) and an LM-driven byte-substitution family. On AWGN, \semOSD\ achieves block error rate (BLER) below the normal-approximation finite-blocklength bound by exploiting source distribution. On a Gilbert--Elliott channel, it reduces BLER by orders of magnitude over Berlekamp--Massey and ordered-statistics decoding (OSD). The latency cost of these gains is one LM forward pass per decoding.}

\bibliographystyle{IEEEtran}
\bibliography{ref/IEEEabrv,ref/OSDAbrv,ref/paper_ref,ref/reference}

\end{document}